\documentclass[a4paper,fleqn]{cas-sc}

\usepackage[authoryear]{natbib}
\usepackage{amsmath}

\def\tsc#1{\csdef{#1}{\textsc{\lowercase{#1}}\xspace}}
\tsc{WGM}
\tsc{QE}
\tsc{EP}
\tsc{PMS}
\tsc{BEC}
\tsc{DE}
\begin{document}
\let\WriteBookmarks\relax
\def\floatpagepagefraction{1}
\def\textpagefraction{.001}
\shorttitle{}
\shortauthors{Bin Gu et~al.}
%\begin{frontmatter}

\title [mode = title]{Bragg's Additivity Rule and Core and Bond model studied by real-time TDDFT electronic stopping simulations: the case of water vapor}   
\author[1,2]{Bin Gu}[orcid=0000-0001-6554-7742]
\author[3,9]{Daniel Mu\~noz-Santiburcio}[orcid=0000-0001-9490-5975]
\author[4]{Fabiana {Da Pieve}}[orcid=0000-0001-6985-9145]
\author[5]{Fabrizio Cleri}[orcid=0000-0003-0272-7441]
\author[3,6,7,8]{Emilio Artacho}[orcid=0000-0001-9357-1547]
\author[9,2]{Jorge Kohanoff}[orcid=0000-0002-8237-7543]
\cormark[1]
\cortext[cor1]{Corresponding author.}
\ead{j.kohanoff@upm.es}

\address[1]{Department of Physics, Nanjing University of Information Science and Technology, Nanjing 210044, China}
\address[2]{Atomistic Simulation Centre, Queen's University Belfast, Belfast BT71NN, Northern Ireland, United Kingdom}
\address[3]{CIC Nanogune BRTA, Tolosa Hiribidea 76, 20018 San Sebastian, Spain}
\address[4]{Royal Belgian Institute for Space Aeronomy, Av Circulaire 3, 1180 Brussels, Belgium}
\address[5]{University of Lille, CNRS UMR8520 IEMN, Institut d’Electronique, Microélectronique et Nanotechnologie, F-59000 Lille, France}
\address[6]{Ikerbasque, Basque Foundation for Science, 48011 Bilbao, Spain}
\address[7]{Donostia International Physics Center DIPC, Paseo Manuel de Lardizabal 4, 20018 San Sebastian, Spain}
\address[8]{Theory of Condensed Matter, Cavendish Laboratory, University of Cambridge, 
J. J. Thomson Ave, Cambridge CB3 0HE, United Kingdom}
\address[9]{Instituto de Fusi\'on Nuclear ``Guillermo Velarde'', Universidad Polit\'ecnica de Madrid, c/Jos\'e Gutierrez Abascal 2, 28006 Madrid, Spain}

\begin{abstract}
The electronic stopping power ($S_e$)
of water vapor (H$_2$O), hydrogen (H$_2$) and oxygen (O$_2$) gases 
for protons in a broad range of energies, centered in the Bragg peak,
was calculated using real-time time-dependent density functional theory (rt-TDDFT) simulations with Gaussian basis sets. This was done for a kinetic energy of incident protons ($E_k$) ranging from  1.56 keV/amu to 1.6  MeV/amu. $S_e$ was calculated as the average over geometrically pre-sampled short ion trajectories. The average $S_e(E_k)$ values were found to rapidly converge with 25-30 pre-sampled, 2 nm-long ion trajectories. The rt-TDDFT $S_e(E_k)$ curves were compared to experimental and SRIM data, and used to validate the Bragg's Additivity Rule (BAR). Discrepancies were analyzed in terms of basis set effects and omitted nuclear stopping at low energies.
At variance with SRIM, we found that BAR is applicable to our rt-TDDFT simulations of $\mathrm{2H_2+O_2}$ $\mathrm{\rightarrow 2H_2O}$ {\it without scaling} for $E_k>40$ keV/amu. The hydrogen and oxygen Core and Bond (CAB) contributions to electronic stopping were calculated and found to be slightly smaller than SRIM values as a result of a red-shift in our rt-TDDFT $S_e(E_k)$ curves and a re-distribution of weights due to some bond contributions being neglected in SRIM. 
\end{abstract}

%\includegraphics{figs/grabs.pdf}

%\item Electronic stopping for protons in H$_2$O, H$_2$ and O$_2$ gasses was calculated via rt-TDDFT.
%\item Averaged stopping converges with few geometrically pre-sampled nm-long trajectories.
%\item Within rt-TDDFT Bragg’s additivity rule was found to be applicable for $E>40$ keV/amu.
%\item Core and bond contributions from rt-TDDFT are consistent with SRIM values.
%\item The electronic stopping power ($S_e$) of water vapor, hydrogen and oxygen gases for protons were calculated with rt-TDDFT simulations. 
%\item With geometrically pre-sampled nm-long ion trajectories, the averaged value of rt-TDDFT $S_e$ converged quickly. The results were compared with observations.
%\item For rt-TDDFT $S_e$, the validity of Bragg's additivity rule (BAR) of $\mathrm{2H_2+O_2}$ $\mathrm{\rightarrow 2H_2O}$ was found to be applicable for $E_k>40$ keV/amu.
%\item The core and bond (CAB) contributions to $S_e$ were calculated from rt-TDDFT simulations, and found smaller than SRIM values.

\begin{keywords}
\sep Electronic stopping power \sep Bragg's Additivity Rule \sep Core and Bond contributions \sep Rt-TDDFT simulation
\end{keywords}
\maketitle

\section{Introduction}\label{intro}
The energy loss of an ion traveling through matter has been a subject of interest since the early days of quantum mechanics \citep{Bragg1905,Rutherford1911,Bohr1913} and it is the starting basis for both detailed calculations and practical applications in ion-beam therapies  \citep{KRAFT2000,Baskar2012,Solovyov2016,friedland2017comprehensive,Durante}, engineering of materials' properties \citep{Calcagno1992,Was2007}, materials in reactors \citep{Granberg2016}, radiation protection for astronauts \citep{ferrari2009cosmic,cucinotta2012,FabianaJGR2021} and induced damage in spacecraft components and on-board equipment \citep{Jiggens2014}. 

The average energy loss rate from the ion to the target material, i.e. the energy transferred per unit distance traveled by the ion, receives the name of stopping power ($S$) and is generally divided into electronic ($S_e$) and nuclear ($S_n$) components as: 
\begin{equation} \label{eq1}
    S=-dE_k/dl=S_e+S_n,
\end{equation}
where $E_k$ is the kinetic energy of the ion and $l$ is the distance travelled by the ion in the target. $S$ is a function of $E_k$ or the velocity of the particle ($v_{p}$). It also depends on the type of projectile and the physical and chemical properties of the target material.
{ Upon slowing down to energies close to the Bragg peak, the ion still moves so fast that} there is no time for the nuclei in the target to react and, hence, nuclear stopping $S_n$ is negligible, while $S_e$ is dominant. As $E_k$ decreases, the importance of elastic scattering from the nuclei and inelastic rotational and vibrational excitation increases according to the incident energy of the ion and properties of the target material~\citep{Cabrera-Trujillo2002a,Cabrera_Trujillo_2020,martinez2020rotational,wang2019collision}. {The latter falls within the adiabatic regime, in which the electrons follow instantaneously the motion of the nuclei, exhibiting a dynamics that is completely tied to that of the nuclei. Adiabatic simulations are useful to describe nuclear stopping, including the possibility of breaking and making of chemical bonds \citep{Kohanoff2017a}. This regime, however, does not include electronic excitations, and hence it is not a useful scenario for describing irradiation with fast projectiles, in the intermediate to high energy region, which is inherently non-adiabatic.}

{The electronic stopping power is a fundamental quantity in the definition of dosimetric quantities such as {\it absorbed dose} and derived operational/protection doses \citep{menzel}. The closely related \textit{restricted} linear energy transfer (LET), defined as energy imparted locally to the medium per unit distance, is a main factor determining the effectiveness of the radiation in inducing biological damage, together with the spatial patterns of energy deposition, the nature of the ion and other chemical and biological aspects  \citep{hall,Paganetti_2014,jchen,schmid}.
Known electronic stopping power tables, such as SRIM~\citep{ziegler2008srim} and PSTAR~\citep{pstar}, are used as pre-initialized values given in input for the low-energy regime to Monte Carlo (MC) condensed history codes for macro- and micro-dosimetric quantities~\citep{ivanchenko2017validation,aposto}, and high-to-low energy datasets are provided as international recommendations, such as the ICRU 49 dataset \citep{ICRU49}, to benchmark track structure calculations in water \citep{FRANCIS20112307,Geant4DNA2018,bern2015}. However, at low energies, SRIM suffers from several limitations and uncertainties \citep{wittmaack2016misconceptions}, such as the use of Lindhard's theory at energies too low for the linear regime to be valid and the use of the Bragg’s additivity rule (BAR) \citep{Thwaites1983}, according to which the stopping power of a compound is the stoichiometric summation of the stopping powers of its constituent elements, neglecting electronic structure and chemical bonding aspects that become relevant at low energies. The ICRU 49 dataset of data \citep{ICRU49} is built such that at low energies (below approximately 0.5 MeV for protons) the tabulated collision stopping powers are based on experimental data and copied directly from the book by Andersen and Ziegler \citep{andersen1977hydrogen} (see also \citep{ziegler1999comments}), on which SRIM is based. Thus, the ICRU 49 dataset actually inherits many of the limitations of SRIM. In particular, the use of the Bragg’s additivity rule at energies around the Bragg peak has shown to overestimate the stopping power by up to 25\% \citep{Thwaites1992}.} In practice, a {\it scaling factor} is often applied to the Bragg curve, so that the scaled curve coincides with available experimental data for the compound, or with calculations at high energies based on Bethe-Bloch theory  \citep{Sigmund2014}. 

To include chemical bonding contributions explicitly, Both et al. \citep{Both1983} proposed to partition the stopping into Core and Bond (CAB) contributions. 
In this scheme, the electronic stopping in compounds is obtained as the superposition of stopping by atomic cores, supplemented with the stopping corresponding to the bonding electrons to consider the ``connectivity''. The core stopping contributions are assumed to follow the BAR, while chemical bonds introduce a correction to core contributions. 
In the SRIM tables, the CAB relative core strengths of a selection of light elements, i.e., H, C, N, O, F, Cl and S, and the strengths of common bonds between them, were determined by fitting the stopping power of light ions (H, He and Li) by a suitable set of molecular targets, using empirical effective charge models for ions \citep{Ziegler2010}. 
It is clear that the CAB scheme used in SRIM relies heavily on simple scaling rules and approximations. To the best of our knowledge, the SRIM CAB $S_e$ values have not been systematically calculated by any other method. 

In the past decade, real-time Time-Dependent Density Functional Theory (rt-TDDFT) started to be applied to electronic stopping power calculations for high-energy ions travelling across a target material, following specific trajectories in nanometer scale samples \citep{ChristopherRace2011,Correa2018,Correa2012,Schleife2015,Yost2017,Ullah2018,Maliyov2018,LI201841,PhysRevA.100.052707}. In this work, we present a rt-TDDFT study of: $(a)$ the $S_e(E_k)$ curves for protons at energies around the Bragg peak {in the paradigmatic case of water vapor}, and H$_2$ and O$_2$ in gas phase, $(b)$ the applicability of BAR to the
%dan reaction 
%dan
{rt-TDDFT obtained stopping values for the case}
$\mathrm{2\,H_2+O_2{\rightarrow}2H_2O}$, and $(c)$ the derived CAB values, comparing them to those reported in SRIM.

For the stopping calculations we successfully applied our previously developed strategy for pre-sampling the projectile's trajectories \citep{Gu2020}, initially tested for liquid water, to the gas phase. Such strategy allows us to select a few short ion trajectories for which the averaged value of $S_e$ calculated by rt-TDDFT reproduces to an excellent extent the variety of different geometrical conditions of the encounters between projectile and target atoms realised in experiments \citep{Gu2020}, thus providing information beyond the "connectivity and conformation" perspective of the CAB approximation. { We chose to focus this study on water vapor for two reasons. Firstly, the available experimental data on water vapor has been, for a long time, the basis for studying energy deposition in liquid water by some first-generation track structure codes \citep{kyri}, and and they still constitute the cross section databases for several well established codes \citep{nikjoo_2006,kyriakou_2022}. Even the codes which have models for liquid water implemented, actually use a mixture of semi-empirical models for liquid and for water vapor. For example, the proton-induced low energy excitations/ionization in Geant4-DNA are based on semi-empirical models parametrized on water vapor data \citep{Miller_1973,Rudd_1985,Rudd_1992,villa}. Dimilarly, processes of electron capture and electron loss by impacting protons and other ions are currently described by analytical parameterizations based on experimental data in the vapor phase \citep{incerti_2018}. Simulations of cross sections in water vapor are also currently performed in (or are the basis of) the recent TILDA-V code \citep{Alcocer-Avila2019}. As currently there are no direct experimental data for excitation and ionization cross-sections for liquid water \citep{kyriakou_2022}, comparisons with existing experimental data in water vapor provide routinely qualitative appreciations of the plausibility of the simulation models. Secondly, the main goal of this study is to understand the validity of additivity rules for computing stopping in water from that in oxygen and hydrogen. By focusing on water vapor we exclude well-documented phase effects present in liquid water due to hydrogen-bonding between water molecules \citep{Bauer_1994}, which complicate the analysis.}

\section{Methods and Simulation details}
\label{sec2}

\subsection{Preparation of water vapor, hydrogen and oxygen gas targets}\label{s2.1}

%---------
\begin{figure}[phtb]
\begin{center}
\includegraphics[width=4.5cm]{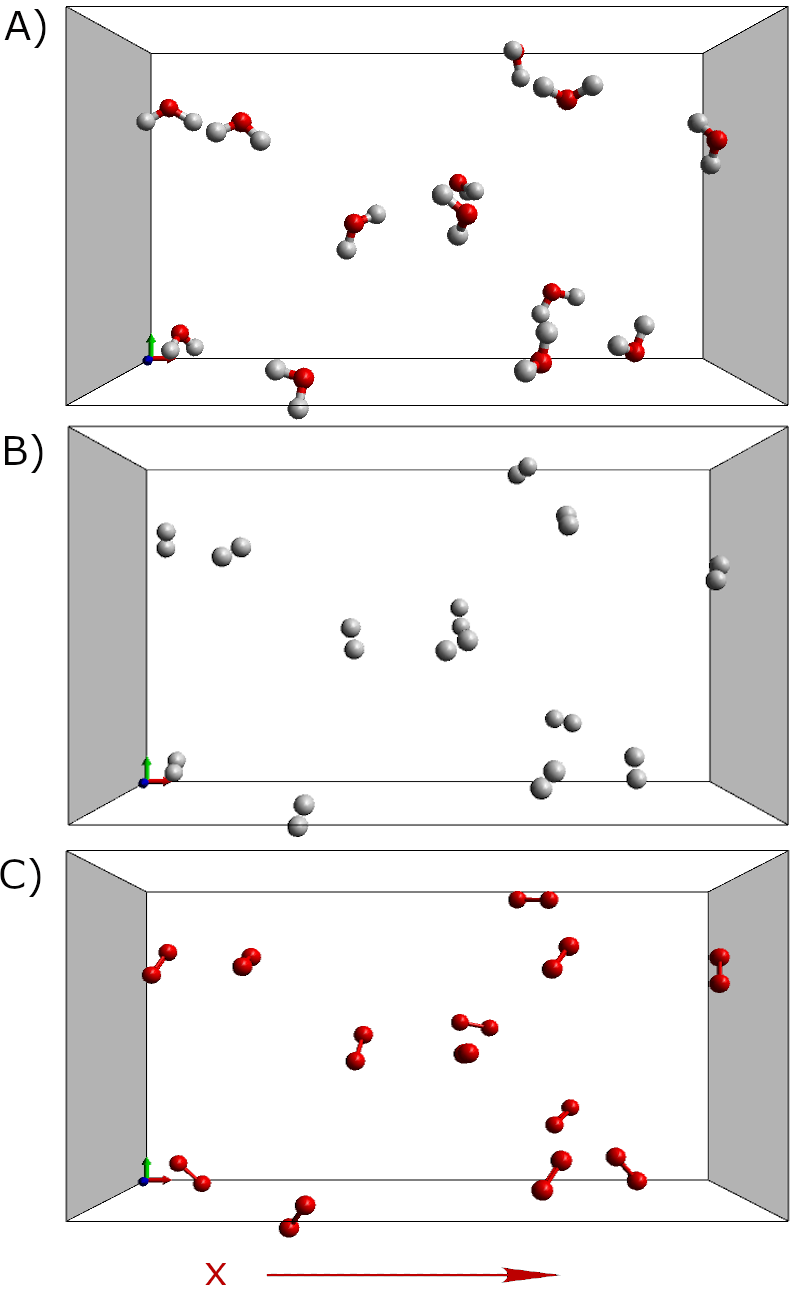}
\end{center}
\caption{The molecular configurations of targets irradiated by swift protons. A) water vapor; B) H$_2$ gas; C) O$_2$ gas. (Color code: oxygen in red; hydrogen in grey.) In each simulation, a swift proton travels through the cell at a given velocity parallel to x-axis. %from pre-sampled initial coordinates. 
}
\label{fig1}
\end{figure}
%---------

All the simulations in this work were carried out with the CP2K code.~\citep{Hutter2014,Kuhne2020}
To create a sample of water vapor, we first carried out a { 25 ps-long} adiabatic \textit{ab initio} molecular dynamics (AIMD) 
%dan simulation \textcolor{purple}{in the Born--Oppenheimer regime} 
simulation (i.e. within the Born-Oppenheimer regime)
of a liquid water sample composed of 104 H$_2$O molecules in the NVT ensemble at $T=300$ K {with a Nos\'e--Hoover chain thermostat}. 
{The electronic structure of the system was treated at the all-electron level
within the GAPW scheme implemented in CP2K. We used the 6-311++G(2d2p) basis set and the Perdew-Burke-Ernzerhof exchange-correlation functional.
Only the $\Gamma$-point in the electronic Brillouin zone was sampled.}
The 104 molecules were placed in a periodic cell of dimensions 21.7$\times$12$\times$12 \AA{}$^3$, resulting in a liquid water density of 0.996 g/cm$^3$. 
Then the density of the sample was reduced to 0.1245 g/cm$^3$ by scaling the coordinates of the center-of-mass of each H$_2$O molecule and duplicating the cell dimensions in the three dimensions. The 13 H$_2$O molecules still contained within the original cell were selected to represent the water vapor target.
The resulting system has a higher density than that of the gas phase at room conditions.
{The reason for studying water vapor with several water molecules in a box is that we can sample several collisions of various impact parameters in a single simulation. Hence, the sampling efficiency is much higher, compared with sampling all possible impact parameters and molecular orientations with one molecule in the simulation box.  An aspect that is not often appreciated is that the cost of periodic-cell simulations, 
%dan2 corrected, not exact (see comment in replyletter)
% which CP2K uses also for isolated molecules, 
which we use here for technical reasons,
grows with the volume of the simulation box. It is therefore more convenient computationally to simulate a smaller box containing more molecules.}

%The resulting system has a higher density than that of the gas phase at room conditions,
%but it has the advantage that including several molecules within the simulation box instead of a single one accelerates the statistical convergence of the stopping power. 
%However, this density is perfectly suitable to produce accurate results for the gas phase as we shall demonstrate in the following.

After volume scaling, the system was reequilibrated for 5 ps
through an additional NVT adiabatic AIMD simulation at $T=300$ K. During this process we fixed the position of the oxygen atoms to avoid collisions between H$_2$O molecules and the formation of hydrogen-bonded water clusters. Finally, we selected the last configuration of this AIMD run (shown in panel A of Fig.~\ref{fig1}) 
for the ensuing rt-TDDFT simulations.   

For convenience, the original samples of hydrogen and oxygen gas were created by replacing each H$_2$O molecule of the water vapor sample with H$_2$ or O$_2$ molecules, respectively. The coordinates of the center-of-mass of each molecule were kept the same. Then, we carried out a {5 ps} AIMD simulation under the same conditions to equilibrate the H$_2$ / O$_2$ molecules, fixing the centers of mass of the molecules.
As before, we selected the last configurations of the trajectories (panels B and C of Fig.~\ref{fig1}) for the rt-TDDFT calculations.

\subsection{Geometric pre-sampling  of short ion trajectories}\label{s2.2}

Since the electronic stopping power is very sensitive to the electronic density, its value depends quite significantly on the specific trajectory of the projectile \citep{Sigmund2014,Correa2018,Dorado1993,pruneda2007,Yao,Gu2020}. 
In order to obtain a meaningful statistically averaged $S_e(E_k)$ that can be compared to {experimental data}, 
it is necessary to run many short trajectories. For targets in condensed phases and randomly selected trajectories, the number of trajectories is on the order of 100 \citep{Yao,Gu2020}. For the gaseous targets studied in this work, it can be expected that more trajectories of similar length are required to achieve an accurate ensemble average of $S_e$, as the electronic density in gaseous targets is distributed more sparsely in the simulation box.

To accelerate the convergence of the running average of $S_e$, we used our ion trajectory pre-sampling protocol based on a geometric criterion, to ensure that the probability distribution functions (PDFs) of the distance from the ion to the closest atoms, $\phi(r_{p{\rightarrow}x})$ (where $x$ is O or H in this work),
accumulated over the selected trajectories, are close enough to the reference PDFs, $\phi_R(r_{p{\rightarrow}x})$, calculated with 50,000 randomly selected trajectories~\citep{Gu2020}. To evaluate the similarity between $\phi(r_{p{\rightarrow}x})$, calculated for of a set of trajectories, and $\phi_R(r_{p{\rightarrow}x})$, we define the overlap index: 
\begin{equation}\label{eq2}
\cap_x=1-0.5\times\int\left|\phi(r_{p{\rightarrow}x})-\phi_R(r_{p{\rightarrow}x})\right|dr_{p}
\end{equation}  
which corresponds to the overlap in the area under the two PDF curves. As the number of selected trajectories (and hence the total accumulated length) increases, all overlap indices approach 1.

For liquid water, a set of 10 short trajectories (about 20 \AA{}-long each) proved sufficient to converge the running average of $S_e$ at the Bragg peak, where the most important dependence of $S_e$ on the trajectory is observed~\citep{Gu2020}. Hence, they can be safely used for carrying out all rt-TDDFT simulations for different ion velocities. 

{Since our simulations were performed under periodic boundary conditions (PBC)
the length of the projectile's trajectories analyzed was contained within a single unit cell in order
to avoid over-excitations produced by the repeated irradiation of the same regions.}
In addition, to prevent the perturbations by excited electrons in neighboring PBC images, a 6 \AA{}-wide vacuum slab was added between neighboring PBC images along the $x$-direction, which is parallel to the projectile's trajectory.

{In each rt-TDDFT simulation, the proton was initially located at the center of the vacuum region and set in motion by applying an instantaneous kick along the $x$-direction}. Then the projectile was allowed to evolve in rectilinear motion for 24.7 \AA{} at the constant given velocity. The portion of the trajectory used to compute $S_e$ (and also the PDFs for trajectory selection) is ${\Delta}L=21.7$ \AA{} from 
{$x = 3$} \AA{} to 24.7 \AA{}, covering the entire target. In addition, possible violent collisions {at the target's entrance and exit points were avoided by discarding candidate trajectories for which the impact parameter at any of these two points was smaller than $1.2$ \AA{}.} Basically, this excludes trajectories that start and/or end with a large variation in the electronic energy ($E_e$ in equation~\ref{eq3}, below), thus avoiding large uncertainties in the calculated value of $S_e$.  {Such close collision events, however, were not discarded when they happened in the central, much larger, part of the trajectory. 
%dan2 rewrite a bit for clarity
An additional concern is that the projectile 
%enters the sample from the vacuum region with the charge in which was initially prepared. 
needs to travel a certain length through the target until its
effective charge stabilizes around a certain steady-state value.
%As the projectile moves through the sample, this charge, e.g. the Hirshfeld charge, evolves until it reaches a stable value and then oscillates around it as it interacts with the molecules in the target. 
We have computed the Hirshfeld charge of the projectile along its path and observed that 
%The present simulations have included this transient in the calculation of $S_e$. Test simulations indicate that discarding the transient region, 
discarding the
transient region prior to reaching the steady state,
which in the vapor phase of water extends to $\approx$ 6 \AA{}, 
%the $S_e$ value changes by 2.5\% at most.
results in a change of $S_e$ of at most 2.5\%,
and hence we consider that such fluctations in the charge state
are irrelevant for the calculation of the stopping power in this case.}

%---------
\begin{figure}[phtb]
\begin{center}
\includegraphics[width=8cm]{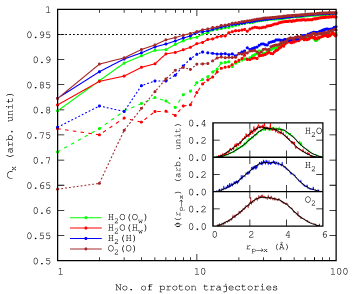} 
\end{center}
\caption{Overlap index $\cap_x$ for increasingly long accumulated trajectories generated by geometric pre-sampling (solid lines with dots) and randomly selected ones (dash-dot lines with dots). The same color code is used for both cases. The dotted line at $\cap_x=0.95$ is shown as a reference for discussion. The inset shows the PDFs using 25 trajectories for geometry pre-sampling, compared to the reference distributions generated by 50,000 randomly selected trajectories (solid black lines). The color code is the same as above.
 }
\label{fig2}
\end{figure}
%---------
As shown in Fig.~\ref{fig2}, for both hydrogen and oxygen gas, only 10 pre-sampled trajectories are required to achieve an overlap index of 0.95, while the same overlap index requires about 70 randomly selected trajectories. For water vapor, a few more pre-sampled trajectories (11 for the oxygen PDF, and 16 for hydrogen one) are required to reach $\cap_x=0.95$. The observation that more trajectories are required for water vapor can be explained by the fact that, for multi-atomic or multi-species targets, the algorithm selects trajectories that optimize the PDFs of two or more species simultaneously. In this case, the trajectories that optimize the oxygen PDF are not necessarily the optimal ones for optimizing the hydrogen PDF, and vice versa. Therefore, more trajectories are needed to achieve convergence for both species.
With random trajectories the overlap index for water vapor is also smaller than for O$_2$ and H$_2$ gas.
The inset of Fig.~\ref{fig2} shows that the reference PDFs of all atoms in the gaseous targets (black smooth lines) can be reproduced extremely well with 25 pre-selected trajectories. 
%------------------------
\subsection{Rt-TDDFT simulation and calculation of  electronic stopping power}\label{s2.3}

%\textcolor{green}{ALL THESE PARAGRAPHS I JUST PUT ON ITALICS ARE `TEXTBOOK' STUFF AND THUS I WOULD ELIMINATE THEM AND SIMPLY CITE SOME NICE BOOK/S OR REVIEW/S -- ALSO I PROPOSE SOME REWRITING OF THE OTHER PARAGRAPHS IN THE SECTION, KEEP THE OLD STUFF COMMENTED} \textcolor{teal}{WHAT ABOUT THIS???} 
%\textcolor{red}{I remove some sentences. The 2018-Alfredo~\citep{Correa2018} is a good review for us. As the ZF scheme is one important topic for this work, it might be worthy to keep the sentences for it.}

Rt-TDDFT, its applications and numerical implementations have been recently reviewed \citep{Ullrich2011,Maitra2016}. The rt-TDDFT approach was used for the first time to compute electronic stopping power in materials by Pruneda et al. \citep{pruneda2007}. %In that seminal work, the ion trajectory was simulated as a moving external potential travelling through the target material. 
Since then, other implementations of rt-TDDFT have been proposed in which the projectile with its basis is explicitly included \citep{Correa2018}. 
When a projectile is forced to move at a constant speed through a target material, the total energy of the system increases by an amount $\Delta{E}$ as a result of the work done by the constrain to maintain the projectile's velocity constant \citep{Schleife2012}. For projectile's kinetic energies large enough, i.e. above a threshold of a few keV/nucleon, the motion of the host nuclei is normally negligible in the time scale of the projectile's transit due to their large mass. Above the energy threshold, there is no appreciable effect on the electronic dynamics if the host nuclei are constrained to stay at their initial positions during rt-TDDFT stopping simulations. The advantage is that, in this way, the change in total energy is due only to the electronic subsystem, $\Delta{E_e}$. The electronic stopping power for a projectile's trajectory of length $\Delta{L}$ can then be calculated as:
\begin{equation}
\label{eq3}
    S_e(E_k)=\Delta{E_e}/\Delta{L}.
\end{equation}

For the calculation of the $S_e$ of protons by gaseous targets we used the all-electron implementation of rt-TDDFT in CP2K~\citep{Hutter2014,Kuhne2020}. 
As in the equilibration and setup of the system, we used the 6-311++G(2d2p) basis set for target atoms, since for water molecules it produces better results than the 6-311G$^{**}$ basis set used in the supplementary information in \citep{Gu2020}. We note that we used the Perdew–Burke–Ernzerhof funcional~\citep{perdew1996generalized} in its time-local version which does not include memory effects (i.e. the exchange-correlation energy and potential are calculated for the instantaneous density using the standard PBE functional).
%
%dan I think here is a nice place to address comment 2.1:
{We remark that the PBE functional has been shown
to provide very good estimations of the electronic stopping power
in water~\citep{Reeves2016,Yao}, and no significant differences have been found
between the regular PBE functional and its hybrid version PBE0~\citep{Reeves2016} or the meta-GGA SCAN functional~\citep{Yao}.}

After placing the proton in its initial position, we obtain the initial Kohn-Sham orbitals, and hence the electronic density, of the system via a regular single-point calculation of the ground state. Then, the rt-TDDFT simulation is initiated by
giving an initial velocity to the projectile.
This velocity, as well as the position of the target atoms, are
maintained unaltered by setting to zero the forces on all nuclei. We called this the Zero Force (ZF) scheme. The time step $\Delta t$ of the real-time propagation was determined by setting a constant displacement of $\Delta x=0.005$ \AA{} in each integration step, i.e. $\Delta t=\Delta x/v$. The largest time step used for our simulations was 0.92 attoseconds, corresponding to the smallest velocity of 0.25 a.u., or energy of 1.56 keV/amu, of the proton. 
%=====================================================
\section{Results and discussion}
\label{sec3}
\subsection{Convergence of rt-TDDFT electronic stopping power with pre-sampled ion trajectories.}
\label{sec3.1}
%---------

%---------
Using the pre-sampled short trajectories selected in Section \ref{s2.2}, we calculated the electronic stopping power $S_e(E_k)$ for protons in water vapor, H$_2$ and O$_2$ gas for proton kinetic energies $E_k$ ranging from 1.56 keV to 1.6 MeV, corresponding to velocities between 0.25 and 8 a.u. This was done by running rt-TDDFT simulations at the theory level and basis set described in Section~\ref{s2.3}. To check the convergence of the averaged $S_e$ values in the gaseous targets, we focused on the case $E_k=125$ keV/amu. This is precisely the proton energy used in SRIM to determine the reference CAB table, which in turn is used to estimate the electronic stopping power for various ions in different compounds~\citep{Ziegler2010}. The rt-TDDFT $S_e$ values at 125 keV/amu will be used later to calculate the rt-TDDFT CAB strengths and compare them with SRIM data. In addition, this energy is close to the Bragg peak, where $S_e$ is most sensitive to the proton trajectory~\citep{Gu2020}.

\begin{figure}[phtb]
\begin{center}
\includegraphics[width=8.5cm]{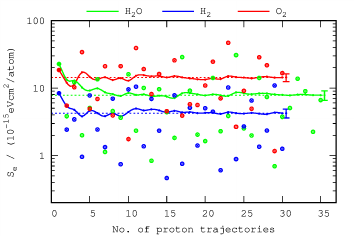}
\end{center}
\caption{Stopping power 125 keV/amu protons in water vapor (green), and H$_2$ (blue) and O$_2$ (red) gas, calculated with 21.7 \AA{}-long pre-sampled trajectories. The individual $S_e$ values for each trajectory are indicated with open circles. The running averages of $S_e$ are shown with solid lines, and standard deviations are shown as a vertical tick to the right of each line. The average $S_e$ values converged over 30 trajectories for H$_2$ and O$_2$ and 35 trajectories for water vapor are indicated with dashed lines.}
\label{fig3}
\end{figure}

In Fig.~\ref{fig3} we show, with open circles, the calculated $S_e$ values for proton at 125 keV/amu along single pre-sampled short trajectories in the three gaseous targets.
A first observation for the three target systems is that the distribution of $S_e$ values of the individual trajectories is quite broad,
with differences that can be larger than a factor of 20 (mind the logarithmic vertical scale in Fig.~\ref{fig3}).
Not unexpectedly, comparing such plot with one for an equivalent condensed phase system~\citep{Gu2020} makes it clear that the
trajectory dependence of $S_e$ is greatly enhanced in the case of gas-phase targets,
which must be taken into account when computing statistically averaged quantities.

{The converged averages of $S_e$ for H$_2$O, H$_2$ and O$_2$ are  $7.8\times10^{-15}$, $4.2\times 10^{-15}$ and $14.3\times10^{-15}$ eVcm$^2$/atom, respectively. They were obtained 
using 35, 30 and 30 pre-sampled trajectories. These values were converged already with
28, 22 and 22 trajectories, remaining stable within $\pm$3\% upon increasing the sampling,
and with final estimated standard deviations of $\pm8\mathrm{\sim}10\%$ as indicated in Fig.~\ref{fig3}.}
%
%dan This is to address a comment by Emilio
{We note that we express $S_e$ in units of eVcm$^2$/atom
as customarily done in the radiation damage field, which has the advantage of being
independent of the density of the target system and thus facilitates the comparison 
with different systems or phase states.}

Simulations conducted at other energies showed that the number of selected short trajectories required for this same level of convergence of $S_e(E_k)$ is less than 30 for water vapor and less than 25 for both oxygen and hydrogen in gas phase. 
However, for the sake of consistency and in order to ensure an utmost statistical accuracy, 
in the following calculations of rt-TDDFT $S_e(E_k)$ curves we imposed a tighter convergence criterion and used 35 pre-selected short trajectories for water vapor and 30 for hydrogen and oxygen gas.   

\subsection{Electronic stopping powers for proton in water vapor, and hydrogen and oxygen gases}
\label{sec3.2}

\begin{figure}[phtb]
\begin{center}
\includegraphics[width=8cm]{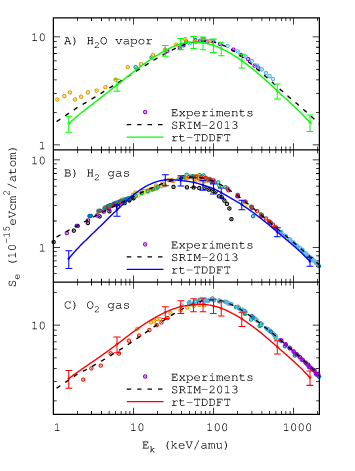} 
\end{center}
\caption{
Electronic stopping power curves for proton by A) water vapor, B) H$_2$ gas and C) O$_2$ gas. The rt-TDDFT curves are show as solid lines with error bars indicating the standard deviation of the averaged values of short ion trajectories.
The corresponding SRIM-2013 curves~\citep{Ziegler2010} are shown as black dashed lines. The experimental values for the corresponding targets are shown as open circles in different colors for the various sources (see details in \citep{IAEA} and references therein).}
\label{fig4}
\end{figure}

The rt-TDDFT $S_e(E_k)$ curves for H$_2$O, H$_2$ and O$_2$ targets calculated using the pre-sampled trajectories are reported in Fig.~\ref{fig4}, along with the corresponding SRIM curves \citep{Ziegler2010} 
and the available experimental data~\citep{IAEA}.
%
%---------
%---------
%
%dan Same paragraph
For water vapor the rt-TDDFT electronic stopping curve
%dan, shown in Fig.~\ref{fig4}A, 
(Fig.~\ref{fig4}A)
coincides with the SRIM curve and with the { experimental data} to an excellent extent, except at very low energies. When compared to SRIM, the position of the Bragg peak in the rt-TDDFT curve is red-shifted by 15 keV/amu, from $E_k=75$ to $60$ keV/amu, and the peak value of $S_e$ is underestimated in $0.34\times\mathrm{10^{-15}eVcm^2/atom}$, i.e. a 4\%.
In a previous work, we calculated the $S_e$ curve for protons in {\it liquid} water, using the (smaller) 6-311G** basis set and found that the position of the Bragg peak in rt-TDDFT $S_e$ curve was also red-shifted by 15 keV/amu, and the peak value of $S_e$ was underestimated by a similar 4\%
~\citep{Gu2020}. 

To assess the contribution of the additional diffuse and polarization functions included in the present basis set, 6-311++G(2d2p), but absent in 6-311G**, we re-calculated the $S_e$ curve for protons in {\it liquid} water with the new basis, {using 16 pre-sampled ion trajectories for the liquid water sample composed of 104 molecules.} 
%dan \textcolor{green}{WHAT ARE THE DETAILS OF THIS? WHICH CONFIGURATION, HOW MANY TRAJECTORIES, ETC?}
The effect of the additional basis functions is to reduce the red shift of the Bragg peak relative to SRIM to 5 keV/amu, and the underestimation of the peak height to %$0.15\times\mathrm{10^{-15}eVcm^2/atom}$, i.e.
less than 2\%. These results indicate that the accuracy of $S_e$ calculated by rt-TDDFT simulations with Gaussian basis sets can be increased by improving the basis, as discussed in the Supplementary Information in~\citep{Gu2020}. However, the absolute quality of the calculated $S_e$ for a given basis set depends on the phase state, whether gas or condensed, and more generally on the density of the target material. 
%dan ??? Hmmm, I would remove the "or density" above, as it raises the question of how much would the density need to change to observe such effect.
Specifically, a larger basis set is required to achieve the same accuracy in water vapor than in liquid water or ice. %{\sout{ More details on the phase effect will be reported and discussed elsewhere.}}
%When $E_k>4$ keV/amu, the values of rt-TDDFT $S_e(E_k)$ are located within the confidence interval of $\pm$8\% of the { experimental data} reported by different groups (see \citep{IAEA} and references therein). 

In the low energy end, when $E_k<5$ keV/amu, the rt-TDDFT stopping values for water vapor fall below the experimental data and the empirical SRIM data, with the relative difference increasing as $E_k$ decreases. The underestimation of rt-TDDFT $S_e$ at low energies becomes even larger when we take into account the red shift due to basis set effects. The red shift can be observed clearly in the high-energy end of Fig.~\ref{fig4}A, where the rt-TDDFT results (solid green line) lie to the left of SRIM (dashed line) and experiment (open circles). This effect has also been reported for other targets in TDDFT~\citep{Schleife2015,Maliyov2018,Gu2020} and quantum Monte Carlo calculations~\citep{Alcocer-Avila2019}. If we shift the rt-TDDFT curve to the right to make it coincide with experiment and SRIM, then the values at low energies become even lower, thus enhancing the underestimation of $S_e$ in that region.

%dan Some changes to make it more clear
%\textcolor{olive}{Firstly, this underestimation of electronic stopping by rt-TDDFT calculations, may have its origin in the possible deviation of the trajectories due to small impact parameter collisions with the target atoms~\citep{daniel}. As the energy loss $\Delta E$ along a trajectory of a deflected ion at low energy will be bigger than the the rt-TDDFT calculation along an rectilinear trajectory with the ZF approximation.}
A reason for this underestimation of the electronic stopping by rt-TDDFT calculations may be the possible deviation of the trajectories due to small impact parameter collisions with the target atoms~\citep{daniel}. Because of these deflections, the projectiles sampled in the experiments after passing a material with a given thickness would have experienced repeated changes of trajectory, thus allowing them to deposit more energy in the material than if the trajectory were rectilinear, as assumed in the ZF approximation. Therefore, we can establish the existence of a threshold energy, $E_{\text{min-ZF}}$, for the validity of the ZF scheme. $E_{\text{min-ZF}}$ should be larger than the energy at which the rt-TDDFT curve and experimental and/or empirical data intersect. According to Fig.~\ref{fig4}A, for water vapor $E_{\text{min-ZF}} > 5$ keV/amu.

In addition, we note that the experimental stopping power for $E_k<100$ keV/amu shown in open circles  in Fig.~\ref{fig4}A, was derived by measuring the ionization range of protons in water vapor applying the continuous slowing-down approximation (CSDA)~\citep{baek2006ionization}. 
The measured stopping power appears larger than the electronic stopping power as the relative contribution of nuclear stopping increases when $E_k$ decreases.
It is worth mentioning that theoretically GEANT4-DNA is suggested to be only suitable for the calculation of stopping and range of protons in liquid water when $E_k>5$ keV/amu, since nuclear stopping is neglected in the program~\citep{francis2011stopping}. 
Clearly, more experimental efforts addressing this important system in this energy range would be welcome. 

For hydrogen gas, the rt-TDDFT value of stopping at the Bragg peak is $6.27\times\mathrm{10^{-15}eVcm^2/atom}$, which is very close to the SRIM value of $6.35\times\mathrm{10^{-15}eVcm^2/atom}$, i.e. within a 1\%. The maximum, however, is located at an energy of 26 keV/amu ($v_p=1.02$ a.u.), which is much smaller than the value of 55 keV/amu ($v_p=1.5$ a.u.) reported by SRIM and { experiments}. As $E_k$ increases the rt-TDDFT $S_e$ curve progressively approaches experimental data and SRIM. On the contrary, as $E_k$ decreases the rt-TDDFT $S_e$ curve decreases much faster than experimental data and SRIM, especially below $E_k<10$ keV/amu. The situation is similar to, but much more dramatic than for water vapor. Also in this case the discrepancy between rt-TDDFT calculations and experimental data can be understood in terms of the limitations imposed by the ZF scheme that does not include trajectories that have scattered from the nuclei.

Therefore, for H$_2$ gas the threshold energy for the validity of the ZF scheme is $E_{\text{min-ZF}} > 10$ keV/amu, quite higher than for water vapor. This is consistent with calculations for a hydrogen beam colliding with molecular hydrogen by Cabrera-Trujillo \textit{et al}, which showed that for projectile energies between 10 and 25 keV/amu, the nuclear and rovibrational contribution of the molecular target introduces an angular dependence on the experimental stopping cross section~\citep{Cabrera-Trujillo2002a}. 
To determine the exact value of $E_{\text{min-ZF}}$, we should go beyond the ZF approximation and disentangle the nuclear and electronic contributions to stopping accurately with a different scheme. This is still a challenge for \textit{ab initio} calculations.

For oxygen gas, the shape of the rt-TDDFT $S_e$ curve as a whole is consistent with the SRIM curve and { experimental data}, especially in the low energy region where the ZF approximation becomes questionable for water vapor and hydrogen gas. 
%dan \textcolor{green}{WHAT HAS THE MASS TO DO WITH ANY OF THIS? THE SENTENCE SHOULD BE REMOVED UNLESS THE RELATIONSHIP IS CLEAR.} 
It is clearly shown that for oxygen gas the threshold energy for the validity of the ZF scheme is lower than for the other two cases, arguably below the lowest energy studied in this work (1.56 keV/amu). We noticed that at the low energy end, the experimental electronic stopping power for protons, shown as red circles in Fig.\ref{fig4}C, was derived by subtracting the  theoretical nuclear stopping powers from the total stopping measured using a differentially pumped stopping cell~\citep{borgesen1985measurements}. Hence those results are substantially lower at low energies compared to values derived from range measurements.
%the Coulombic repulsive force on a close impacting ion by the O$_2$ molecule is larger than that by both water and H$_2$, and the mass of the O$_2$ molecule is larger.
%\textcolor{green}{[The previous "the" refers to which results??? to the ones by Borgensen 1985??? In such case put "those" and not "the"]}
%
%
%dan \textcolor{red}{NOTE: As pointed out by Daniel, there is some logical problems. I think it is better to delete the arguments about mass effects and Coulomb force.}
%dan \textcolor{teal}{HMMM, DON'T QUITE UNDERSTAND THE ARGUMENT STILL. MORE CHARGE AND MORE MASS ON THE O MEANS THAT THE PROTON IS MORE DEFLECTED, I.E. BIGGER CHANGE IN TRAJECTORY, THUS ZF SHOULD BE WORSE. AS FOR THE CHANGE IN ABSOLUTE VALUE OF SPEED, I DON'T SEE SO CLEAR THAT THE SPEED IS LESS CHANGED, THAT WOULD DEPEND ON MOMENTUM CONSERVATION AND I DON'T SEE THAT CLEARLY THAT THE MOMENTUM OF THE PROTON IS LESS CHANGED.}
%
%
%
In addition, while the shape of the rt-TDDFT $S_e$ curve is well reproduced, there is a noticeable red-shift of the whole curve; the rt-TDDFT Bragg peak for protons in oxygen gas is $15.58\times\mathrm{10^{-15}eVcm^2/atom}$, which is lower than the SRIM value of $17.15\times\mathrm{10^{-15}eVcm^2/atom}$ by 9\%. The position of the rt-TDDFT Bragg peak is red-shifted by 25 keV/amu from 100 keV/amu ($v_p=2.0$ a.u.) of SRIM  to 75 keV/amu ($v_p=1.73$ a.u.) of the rt-TDDFT calculation.
As pointed out above,
we have already established that this red shift is mainly related to basis set convergence.

In this work, the number density of molecules of the three targets are set to be equal. Therefore, the ratio of the number density of electrons, $\rho_e$, is 16:10:2 for O$_2$, H$_2$O and H$_2$, respectively.
As can be seen in Fig.~\ref{fig4}, at high energy, where the ZF scheme is perfectly suitable, the magnitude of the red-shift, $\delta{E_k}=[E_{k(\mathrm{ref})}(S_e)-E_{k(\mathrm{rt-TDDFT})}(S_e)]$ for a given $S_e$ value, increases with $\rho_e$ according to $\delta{E_k}[\mathrm{H}_2]<\delta{E_k}[\mathrm{H}_2\mathrm{O}]<\delta{E_k}[\mathrm{O}_2]$. 
The fact that $\delta(E_k)$ is positively correlated with $\rho_e$ constitutes further evidence that the red-shift is related to the completeness of the basis set.
{In addition, these results suggest that the basis set for hydrogen seems closer to completeness than that for oxygen.}

It should be mentioned that the rt-TDDFT $S_e$ curves in Fig.~\ref{fig4} are all calculated based on the restricted Kohn-Sham (RKS) approximation~\citep{kohanoff2006}. This means the channels for spin flipping and related electronic excitations have been neglected. 
%dan This is dangerous, the reader could think that the results for O2 are bullshit. I try to reword in a more positive way: 
%Actually, for oxygen gas one should use the unrestricted Kohn-Sham (UKS) scheme, as the ground state of the O$_2$ molecules is a spin triplet. This is possible, and we have done it for comparison with the RKS simulations, obtaining similar results, 
{We note that we tested the effect of using the unrestricted Kohn-Sham (UKS) scheme for the case of the O$_2$ molecule
(which actually could be thought to be imperative since such formalism is required to obtain the triplet ground state of O$_2$),
finding out that both RKS and UKS produce quite similar $S_e$ values, which justifies the use of RKS over UKS given the smaller computational cost for the already expensive determination of S$_e$ via rt-TDDFT.}
In any case, a suitable scheme for the propagation of the time-dependent Kohn-Sham orbitals should allow for spin-flipping channels. In the present calculations this was not implemented and, hence, any such excitations were absent. This may explain the underestimation of $S_e$ for O$_2$ gas. Simulation techniques to incorporate these excitations are still under development~\citep{Kuhne2020,casanova2020spin}. 

%dan XC issue treated in methods section.
%{xxxx exchange correlation xxxx}

\subsection{Bragg's additivity rule: from hydrogen and oxygen to water} 
\label{sec3.3}

The assumption of BAR is that the stopping power in a compound can be approximately calculated as the stoichiometric sum of the stopping powers of its elementary components:
\begin{equation}
   S_e^\mathrm{BAR}(E_k)=\sum {r_\mathrm{x}}~{S}_{e}{[\mathrm{x}]}(E_k),
   \label{bar}
\end{equation}
in which $r_\mathrm{x}$ is the ratio of the number density of the atomic species $\mathrm{x}$ to the total number density of all atoms in the target. ${S}_{e}{[\mathrm{x}]}(E_k)$ is the stopping power in the elemental material of species $\mathrm{x}$. This expression carries the implicit assumption that the 
physical phase (e.g. solid vs liquid) of the stopping medium and the chemical bonding of atoms into molecules have negligible influence on the mean energy loss.

However, differences in electronic structure in going from free atoms to molecules cannot be neglected, especially in the low energy regime, where the relative contribution of valence (outer shell or bonding) electrons to the stopping power is large. This is also the case of light elements whose valence electrons are a major fraction of the total. Under those circumstances,  the validity of the BAR becomes questionable~\citep{Thwaites1983}. 
A measure of the bonding contribution is given by the difference in (measured or calculated), stopping power \textit{relative} to the BAR approximation, i.e.
\begin{equation}\label{eq-seb}
\Gamma(E_k)=[1-S_e(E_k)/S_e^\mathrm{BAR}(E_k)]\times 100\%.
\end{equation}

In practice, the concept of \textit{elementary} material, as used by SRIM, refers to stable forms of the element in the gas phase, which in many cases corresponds to simple molecules instead of isolated atoms. For H$_2$O, they are O$_2$ and H$_2$. Therefore, the bonding contribution in equation (\ref{eq-seb}) is the \textit{relative} variation of $S_e$ as the target changes from a mixture of the elementary molecular species, to the compound. 

%---------
\begin{figure}[phtb]
\begin{center}
\includegraphics[width=8cm]{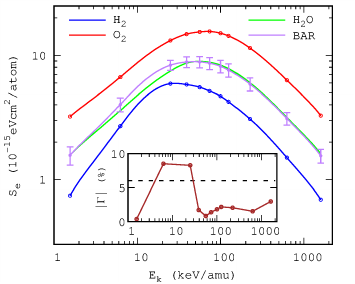} 
\end{center}
\caption{The BAR additivity of rt-TDDFT electronic stopping power $S_e(E_k)$, according to formula of $\mathrm{2H_2+O_2{\rightarrow}2H_2O}$. The rt-TDDFT $S_e(E_k)$ curves for H$_2$ gas and O$_2$ gas, and water vapor are shown with blue, red and green lines individually. The BAR curve is given as purple line with error bar. The inset shows the relative scaling factor $\Gamma$ obtained from rt-TDDFT (brown dots and line), along with the constant scaling value recommended by SRIM (black dashed line).}
\label{fig5}
\end{figure}
%---------

We used the results of Fig.~\ref{fig4} to assess the validity of Bragg's additivity for the electronic stopping of water calculated by rt-TDDFT, where the mixing is decribed by the reaction $\mathrm{2H_2+O_2{\rightarrow}2H_2O}$. The resulting curves are reported in Fig.~\ref{fig5}. The error bars are calculated as the sum of those for O$_2$ and H$_2$ (see Fig.~\ref{fig4}). As shown in the inset to Fig.~\ref{fig5}, when $E_k>40$ keV/amu ($v_p>1.25$ a.u), the relative scaling factor $\Gamma$ stays below $\pm$3\%, which is below the statistical error of rt-TDDFT $S_{eBAR}$.  This means that BAR is supported by parameter-free rt-TDDFT calculations {\it without further scaling}, in the medium- to high-energy regime, including the Bragg peak at 72 keV/amu. When $E_k$ decreases below 40 keV/amu, $S_{eBAR}$ is larger than the $S_e$ calculated directly. $\Gamma(E_k)$ first increases to $\sim 8.5$\% for $E_k=20$ keV/amu and $E_k=5$ keV/amu, and then decreases to almost zero on the low energy end of $E_k=1.56$ keV/amu.

It is important to remark that the energy dependence of $\Gamma(E_k)$ is quite different from that used in SRIM. In the latter, from H$_2$ and O$_2$ to water, $\Gamma$ is assumed to be 6\%, independently of the projectile's energy. This value is obtained by scaling the BAR curve to reproduce { experimental data}, if available, or otherwise the CAB value for a compound at the Bragg peak, and then it is assumed to be valid at all velocities. Our results indicate that this may not be the most suitable approximation to $\Gamma$.

It has been mentioned in Section~\ref{sec3.2} that, on the low-energy end, the rt-TDDFT $S_e$ calculated within the ZF approximation is underestimated with respect to { experiments}. This makes difficult the comparison of the calculated $\Gamma(E_k)$ with that arising from experimental data. Nevertheless, the decrease of $\Gamma(E_k)$, which is inconsistent with the notion that the chemical bonding contribution to electronic stopping should emerge clearly in the low-energy regime, implies that electron-nuclear coupling is important for light targets and low velocities. 

\subsection{Core and bond (CAB) electronic stopping powers}
\label{sec3.4} 
The SRIM CAB table was calculated according to the following prescription \citep{Ziegler2010,ziegler2008srim}. 
Firstly, the {experimental} data points available for different ions at various velocities and in different materials were scaled to the equivalent value corresponding to protons at 125 keV/amu, assuming that the effective charge of the ion does not depend on the target material, but only the energy of the ion. Next, the core and bond contributions for protons at 125 keV/amu were disentangled by solving the system of linear equations
\begin{equation}   \label{eqcab}
   S_e^\mathrm{CAB}(E_k)=\sum{S_e[X]}+\sum{S_e[Y]},
\end{equation} 
in which [X] indicates the ionic core, i.e. nuclei plus core electrons and [Y] refers to the inter-atomic chemical bonds in the target material, involving only valence electrons. For example, for a CO molecule $S_e^\mathrm{CAB}=S_e[$C$] + S_e[$O$]+S_e[$C$\equiv$O$]$. Finally, the whole stopping power curve as a function of the energy (or velocity) of the ion was generated by scaling the BAR curve with the ratio of ${S_e}/{S_e}^\mathrm{BAR}$ for protons at 125 keV/amu, and multiplying with the following asymptotic function
\begin{equation}
f(v_1)=\left\{1+\exp \left[1.47\left(\frac{v_{1}}{v_{0}}-7.0\right)\right]\right\}^{-1},
\end{equation}
with $v_0=1$ a.u., corresponding to 25 keV/amu, to ensure that the deviation from the BAR curve vanishes at high energies~\citep{ziegler1988stopping}.

Based on rt-TDDFT calculations of $S_e$ for protons at 125 keV/amu in different compounds, the CAB contributions to electronic stopping powers can be directly disentangled by solving a system of linear equations, without resorting to a general fitting as it is done in SRIM. Here we calculated the rt-TDDFT core and bond contributions of  hydrogen and oxygen in water, and compared them to SRIM published tables.

In addition to water vapor and H$_2$ and O$_2$ gases, we included also a  calculation of the stopping power for protons in hydrogen peroxide (H$_2$O$_2$), with the same procedure, theory level and basis set used for O$_2$ and H$_2$ gases, as discussed in Section 2.
This introduces a different type of bond between two oxygen atoms (O-O), weaker than that in O$_2$ (O=O). 
On the other hand, since the hydrogen electron always participates in a chemical bond, it is reasonable to follow the rule used in SRIM, i.e. taking the core contribution of the hydrogen atom $S_e[\mathrm{H}]$ as zero.
According to Eq. (\ref{eqcab}), 
the rt-TDDFT stopping values for H$_2$, O$_2$, H$_2$O and H$_2$O$_2$ can be combined in the following set of four linear equations in five unknowns: 
\begin{equation}
\begin{aligned}
\begin{cases}
  	S_e[\mathrm{H_2}]
	=S_e[\mathrm{H\text -H}]=8.46\pm 0.61\\
	S_e[\mathrm{O_2}]
	=2S_e[\mathrm{O}]+S_e[\mathrm{O\text=O}]= 28.68\pm 2.13\\
    S_e[\mathrm{H_2O}]
    =S_e[\mathrm{O}]+2S_e[\mathrm{O\text-H}]= 23.41\pm 1.90\\
    S_e[\mathrm{H_2O_2}]
    =2S_e[\mathrm{O}]+2S_e[\mathrm{O\text-H}]+S_e[\mathrm{O\text-O}]=33.55\pm3.13 
     \end{cases}
   % S_e(\mathrm{H_2O_3})
   % =3S_e[\mathrm{O}]+2S_e[\mathrm{O\text-H}]+2S_e[\mathrm{O\text-O}]=49.73\pm3.58, 
\end{aligned}
\label{eq7}
\end{equation}
%in which the unit of stopping power is 
where the stopping power is expressed in
$\mathrm{10^{-15}eVcm^2/unit}$, 
being `unit' an atom, bond, or molecule. This set of equations has an infinite line of possible solutions.  
In order to obtain a unique solution,
we fixed the ratio $S_e[\mathrm{O}]/S_e[\mathrm{O\text=O}]$ to the SRIM value, thus reducing the system to four equations in four unknowns. %which has a unique solution. 
The calculated uncertainties are also decomposed into core and bond contributions according to their stoichiometric coefficients in Eq. (\ref{eq7}).

\begin{table}[htbp]       
\centering
\caption{Hydrogen and oxygen related CAB contributions to electronic stopping power calculated by rt-TDDFT simulations and SRIM-2013~\citep{Ziegler2010}. The unit of stopping power is $\mathrm{10^{-15}eVcm^2/(atom~or~ bond).}$ The last column reports the relative percentual difference between rt-TDDFT and SRIM. 
}
\begin{tabular}{ccccc}
\toprule
&Name& rt-TDDFT & SRIM-2013 & $\epsilon(\%)$ \\
%\hline
% \multirow{2}{*}{Core}   & H  & 0  &       0  & 0 \\
%                         & O  &4.853$\pm0.35$ & 5.446 & -10.9 \\
%\hline
% \multirow{4}{*}{Bond}&H-H  &8.46$\pm0.61$   & 9.590  & -11.8 \\
%            &H-O  & $9.278\pm0.77$  & 8.758 & ~5.94  \\
%            &O-O  &  $5.286\pm$1.27 &  --  & -- \\
%            &O=O  &18.973$\pm1.43$ & 21.290 & -10.9 \\
\midrule
 \multirow{2}{*}{Core}   & H  & 0  &       0  & 0 \\
                         & O  &4.68$\pm0.35$ & 5.36 & -12.6 \\
\midrule
 \multirow{4}{*}{Bond}&H-H  &8.46$\pm0.61$   &10.049  & -15.8 \\
            &H-O  & $9.37\pm0.77$  &10.085 & -7.1   \\
            &O-O  &5.$1\pm$1.27 &  --  & -- \\
            &O=O  &19.24$\pm1.43$ & 22.044 & -12.6 \\
\bottomrule
\end{tabular}
\label{tb1}
\end{table}
%-------

The CAB contributions to the calculated rt-TDDFT stoppings, along with the SRIM-2013 data, are presented in Table~\ref{tb1}.
On the other hand, as shown in Fig.~\ref{fig2}, at $E_k=125~\mathrm{KeV/amu}$ all the values of rt-TDDFT $S_e$ are underestimated relative to SRIM.
Hence, all the rt-TDDFT CAB values are smaller than the corresponding SRIM data by more than 10\%, except the O-H bond which is only about 7\% smaller, 
as a consequence of the calculated $S_e$ for H$_2$O being the closest one to SRIM (Fig.~\ref{fig2}).

For O$_2$ the discrepancies between rt-TDDFT results and { experimental data} are due to the basis set effect discussed above, which induces a red-shift in the rt-TDDFT curve. For H$_2$, in that energy region the ZF approximation is probably valid, but there is an obvious discrepancy between different sets of experimental data. SRIM follows one set, which runs slightly above rt-TDDFT calculations, and this may contribute to the underestimation.

It should also be noticed that, in the region of the Bragg peak, stopping powers for protons and $\alpha$-particles in water and organic materials in the vapor phase are often larger than in the liquid or solid state by up to 5-10\%. The phase effect in $S_e$ is considered to be due largely to changes in electronic excitation levels across the phase transition \citep{Thwaites1992}. {For liquid water and water vapor, the difference arises from the presence or absence of hydrogen bonds between the water molecules. In practice, by scaling up the size of the box from liquid water, the O$\cdots$H distances become larger than 4 \AA{}, well longer than the typical 1.8--2.6 \AA, so that hydrogen-bonding interactions  are negligible. Therefore, we argue that the vapor state is very well-represented in  the  current  approach  and  we  consider that the error associated to using such density is negligible. To substantiate this assertion, we have run tests at a lower density, and observed no significant differences.}
%So that the vapor state has been represented very well.}
%But in principle, simulations at a lower number density would be required to assess the influence of density effects on the discrepancies between rt-TDDFT and SRIM in the CAB contributions.}

Interestingly, for both rt-TDDFT calculations and SRIM, the magnitude of the bond contributions reported in Table~\ref{tb1} follows the order:  $S_e[\text{O=O}]>S_e[\text{H-O}]>S_e[\text{H-H}]>S_e[\text{O-O}]$. Notice that $S_e[\text{O-O}]$ is not provided by SRIM, though. This order is consistent with that of the average bond energies for these bonds \citep{luo2007}. For example, the  O-O bond strength is 142 kJ/mol and its average rt-TDDFT $S_e$ value is quite small, at $5.1\times\mathrm{10^{-15}eVcm^2/bond}$. In contrast, the bond energy of the O=O double bond, the strongest one in the table, is 494 kJ/mol, which correlates with a larger contribution to stopping power, i.e. $19.24\times\mathrm{10^{-15}eVcm^2/bond}$ according to rt-TDDFT simulations. This correlation suggests that the bond contribution to the electronic stopping power is proportional to chemical bond strength, which is reasonable as stopping is known to increase with electronic density. This observation is consistent with more elaborate addition schemes in which the stopping of the compound depends on the number of valence electrons \citep{Sigmund2018}.

It is important to remark that this decomposition depends heavily on the bonds included in the fit. For example, if we ignore the O-O bond by setting $S_e[\text{O-O}]=0$, while keeping the last equation in (\ref{eq7}), the $S_e[\text{O}]$ core contribution increases to compensate for the missing bond term. If we set the unrealistic condition that  $S_e[\text{O-O}]=S_e[\text{O=O}]$, then the other core and bond contributions decrease. One could also ignore the equation for H$_2$O$_2$ in (\ref{eq7}). In that case the core and bond contributions are similar to the original ones, except that there is no bond contribution for O-O.

%=====================
\section{Conclusions and perspectives}
 
Rt-TDDFT simulations are by now a well-established tool to compute electronic stopping in complex systems directly, when experimental data are not available or incomplete. Moreover, they can be used to calculate the stopping for elements or simple molecules, thus forming a data base to compute stopping power in general complex systems via additivity rules. 

In this work, electronic stopping power curves for energetic protons in water vapor, hydrogen and oxygen gas, were calculated via rt-TDDFT simulations, as a function of the kinetic energy of the proton. The approach used the zero force approximation, and a recently developed geometric pre-sampling technique to select the projectile's trajectories.
We found that, for these gaseous targets, a number of 25-30 optimally selected, 2 nm-long ion trajectories, are required 
to converge the average of $S_e(E_k)$ within a 3\%. The obtained values can be compared directly with experimental data. 

For water vapor and hydrogen gas, in the high-energy region, the rt-TDDFT Bragg peak value and the shape of $S_e(E_k)$ curve are consistent with { experimental data}. On the other hand, at low energies, below a threshold of 5 keV/amu for water vapor and $\sim20$ keV/amu for H$_2$ gas, rt-TDDFT stopping is underestimated compared to experimental data. The discrepancies are consistent with the introduction of the zero force scheme, which neglects scattering of low velocity ions by nuclei. This is especially important for light atoms like H, which are present in H$_2$ and H$_2$O, but not in O$_2$. 
For O$_2$ gas, there is a clear red shift of the rt-TDDFT $S_e(E_k)$ curve compared to experimental data, much more than for water vapor and H$_2$ gas. We ascribe this to basis set convergence effects. %Further studies with optimized basis set \citep{Maliyov2020} should be carried out to improve the accuracy.

Bragg's additivity rule for the  $\mathrm{2H_2+O_2{\rightarrow}2\,H_2O}$ system was found to be applicable to rt-TDDFT electronic stopping values without scaling, when $E_k>40$ keV/amu. Our results indicate that the scaling factor depends on velocity, hence suggesting that the constant 6\% scaling proposed by SRIM for water may not be the most suitable approximation. The rt-TDDFT core and bond electronic stopping contributions are smaller than those derived from SRIM data, as a result of the red shift of rt-TDDFT $S_e(E_k)$ curves, but the relative weights of the core and bond contributions are consistent between the two methodologies, thus further cross-validating the approach.

Therefore, to increase the general accuracy of rt-TDDFT calculations of electronic stopping power with high efficiency, it is worth focusing on the following two aspects. In the first place, the red-shift of the $S_e$ curves might be mitigated by increasing the basis set size, by designing optimal Gaussian basis sets for this type of application~\citep{Maliyov2020}, or by using plane wave methods and sufficiently high cutoffs \citep{Correa2018}, possibly via GPU implementations~\citep{andrade2021inq}. Secondly, for light targets and low ion velocity regime, one should move away from the zero force approximation and allow for the target atoms to move, e.g. via Ehrenfest dynamics. This will most likely require a modification of the geometric sampling algorithm, which should be designed to include corrections for non-rectilinear trajectories ~\citep{Cabrera-Trujillo2002a}.

The SRIM CAB table for light compounds includes the elements: H, C, N, O, F, S, and Cl, and was fitted to experimental data on 114 different compounds \citep{Ziegler2010}. To move into that direction we are carrying out additional rt-TDDFT simulations for a set of relatively simple molecules including the other elements, e.g. CO$_2$, CH$_4$, \textit{etc.}

In addition to additivity of stopping power from elementary atoms to molecules, simple organic molecules can also be used as units to calculate stopping power for complex targets such as DNA and other biomolecules like proteins \citep{Sauer2019}. In this case, instead of core and bond, we would have a superposition of intrinsic contributions from individual molecules, and contributions from bonds between molecular units. These can be covalent or hydrogen bonds.

\section*{Declaration of competing interest}
The authors declare that they have no known competing financial
interests or personal relationships that could have appeared to influence the work reported in this paper.

\section*{Acknowledgements}
This work has received funding from the Research Executive Agency under the EU's Horizon 2020 Research and Innovation program ESC2RAD (grant ID 776410). We are grateful for computational support from the UK national high performance computing service, ARCHER, for which access was obtained via the UKCP consortium and funded by EPSRC grant ref EP/P022561/1. This work benefited from networking activities carried out within the EU funded COST Action TUMIEE (CA17126) and represents a contribution to it.


\bibliographystyle{cas-model2-names}

%\bibliography{watervapor}

\end{document}